\documentclass[final,1p,times]{elsarticle} 
\usepackage{graphicx} 
\usepackage{amssymb} 
\usepackage{amsthm} 
\usepackage{lineno} 
\usepackage{epsfig}
\journal{Nuclear Physics A} 
\begin{document} 

\begin{frontmatter} 


\title{The effect of quark coalescence on conical signals}

\author{Vincenzo Greco$^1$, Giorgio Torrieri$^2$,Jorge Noronha$^3$,Miklos Gyulassy$^3$}

\address{$^1$ Dipartimento di Fisica e Astronomia, Universita di Catania, Italy and  INFN-LNS, Laboratori Nazionali del Sud, Italy
}

\address{$^2$ FIAS, Johann Wolfgang Goethe
Universit\"{a}t}

\address{$^3$ Department of
Physics, Columbia University, 538 West 120$^{th}$ Street, New York,
NY 10027, USA}

\begin{abstract} 

We explore the effect of hadronization by partonic coalescence on a ``conical'' signal at the partonic level.  We show that, by transferring partons from a lower to a higher $p_T$, coalescence makes the conical signal stronger and hence less susceptible to thermal smearing, provided the signal is integrated over a large momentum bin and effects such as non-collinearity and a finite Wigner function width are taken into account.  We explore the role of this effect in baryon/meson scaling and calculate the effect of resonances decays on such a conical signal.
\end{abstract} 

\end{frontmatter} 



An experimental observation that has aroused quite a lot of recent theoretical interest in the heavy ion community is the detection, in hard-soft particle correlations, of a pattern commonly associated with ``conical flow'', \cite{expstar,expphenix}.  For two-particle hard-soft correlations, it was found that the soft away-side bump, once the flow contribution has been subtracted, exhibits a two peak structure.
Such a correlation could, in principle, arise if the energy-momentum deposited by the jet is thermalized and released into sound waves.  As was known from the last century, such waves produce a characteristic conical interference pattern, whose angle is related to the speed of sound \cite{lifshitzlandau}.    Finding such a pattern in heavy ion collisions \cite{stoecker,shuryakmach} would confirm the general consensus that the matter created in heavy ion collisions behaves like a ``perfect fluid'' \cite{whitebrahms,whitephobos,whitestar,whitephenix} (though the signal has been observed at energies where flow observables are well below the perfect fluid limit \cite{szuba}) and provide a way to link two-particle correlations directly to the equation of state.

Unfortunately, while the conclusive detection of Mach cones would provide evidence for the perfect fluid behavior, the opposite is not the case.
Many hypothetical situations exist where, while the fluid is perfect, the Mach cone will not be observed at the end of its evolution.  As shown in \cite{school,prl,betzqm,betz1,betz2,heinzmach}, two factors strongly go against a conical correlation even in a {\em perfect} fluid:
Firstly, The thermal broadening in a Cooper-Frye type freeze-out \cite{cf}, which, to first order in $p_T U/T$ (where $p_T$ is the transverse momentum of the produced parton, $U$ is the flow component magnitude and $T$ is the temperature) just gives a phenomenologically uninteresting ``broad away-side peak''.  Hence, to see a cone, one would either have to be in the non-linearized hydrodynamic regime (where what is being seen is not a ``true'' Mach cone \cite{prl,betzqm,betz1,betz2,heinzmach}), or at high $p_T$ (where the effect of non-thermalized partons can not anymore be neglected).
The second factor is the formation of a ``diffusion wake'' if the parton deposits momentum as well as energy in the medium.   The wake contributes to the broad away-side peak.

The fact that Mach cones are not an inevitable consequence of the ``hydrodynamics+Cooper-Frye'' paradigm, together with their experimental detection, suggests that a hadronization mechanism different from Cooper-Frye freeze-out might have a role in creating the conical signal.
That possibility is, in fact, hinted at by the momentum range where Mach cone signals are seen \cite{expphenix}:  At the moment, the consensus is that up to a momentum of 1-1.5 GeV hydrodynamics with a Cooper-Frye freeze-out provides a good description of soft data.  At higher momentum, non-equilibrium effects such as quark coalescence \cite{molnarpaper,muellerpaper,hwapaper,grecopaper} may appear.
\cite{expphenix} shows that the away-side trigger required for a robust ``conical'' signal goes well into the coalescence regime, and in fact the away-side signal changes very little between the hydrodynamic and the coalescence regimes.
(Note that, if coalescence happens by kinetic energy and not momentum, coalescence can fit elliptic flow ($v_2$) data until a negligible transverse momentum \cite{lacey}).
Thus, one naturally {\em expects} that coalescence might play a big role in the generation of conical-like signals, and it is worth investigating what this role might be.  This is the purpose of the current work.

\begin{figure}[h]
\begin{center}
\epsfig{width=11cm,figure=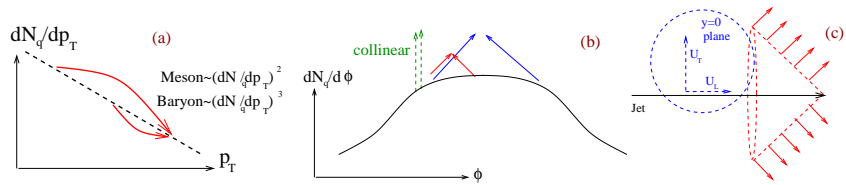}
\caption{\label{gausspeak}(c):The coordinate system used for the conical flow ansatz. The other panels show the qualitative illustration of possible coalescence effects in the $\phi$ (middle (b) panel) and $p_T$ (left (a) panel) plane}
\end{center}
\end{figure}

It is immediately clear that the effect of the simplest collinear coalescence model (the one associated with the $v_2(p_T^h)=n v_2(p_T^q/n)$ scaling of mesons and baryons \cite{molnarpaper}) is to amplify an already existing conical signal (by a factor of 2 for mesons and 3 for baryons) rather than create one (Fig. \ref{gausspeak} (a)).  
 Generally it can be shown that coalescence can enhance locally the angular correlation and from
a quark distribution without a double peak structure it can generate two peaks. On the other hand
if the quark distribution is exponential (as is the case for a locally thermalized system freezing via Cooper-Frye) in the entire $p_T$ range, the weakening of the signal at low $p_T$ is balanced with the enhancement due to the shift to the higher hadronic $p_T$ (Fig \ref{gausspeak}(a)), so the hadron distribution from coalescence and Cooper-Frye become very similar.

This scaling is broken in non-collinear coalescence (expected from energy-momentum conservation when quark masses are non-negligible, eg in constituent quark coalescence), where the angular distribution has the potential to change (Fig. \ref{gausspeak} (b)) .  This effect is amplified if one assumes a local, but not $\delta-$function Wigner function \cite{muellerpaper}, so nearby, but not perfectly overlapping quarks (ie, quarks emitted from cells flowing in slightly different directions) can coalesce.  For the subsequent calculations, we have used the coalescence Montecarlo, incorporating these features, developed in \cite{grecopaper}.   The parameters (constituent quark masses, Wigner function etc.) are tuned to reproduce baryon/meson and $v_2$ data \cite{grecopaper}.    
It should be underlined that the effects described here depend on such a ``realistic'' coalescence model.  If coalescence is perfectly collinear, the distributions of the locally thermalized partons, are locally much closer to exponentials, at the weakness of the signal at lower $p_T$ exactly cancels out the sharpening of the signal by coalescence.
The quark distribution function we used exhibits a generic ``conical flow'' in a static medium (transverse or elliptic flow of the background are neglected), expressed in  the coordinate system, illustrated in Fig. \ref{gausspeak} (c).
``longitudinal'' ($L$) and transverse ($T$) directions are defined in terms of the {\em jet} and are both transverse with respect to the beam
(rapidity=0). In this coordinate system, the general quark distribution function with a constant thermal background and a generalized
``conical'' flow can be well parameterized by the following Ansatz 
\begin{equation} 
f_q(p,\phi,\phi_r)=\rm
Exp\left[-\frac{  (p^2+m^2)^{1/2} U_0 + p U_L cos\phi + p U_T sin\phi sin\phi_r }{T} \right] \label{Eq.2} 
\end{equation} 
where $p$ and $m$ are the quark momentum and mass, $U_0,U_L,U_T$ are the flow components, and $\phi_{r},\phi$ are the angles, respectively, in configuration and momentum space.

This ansatz
coincides with the asymptotic Mach cone limit examined in \cite{prl} and shown to generate one away-side peak (rather than a cone) to first
order in $p_T U_0/T$.   It also leads to diffusion wake dominance if $U_L>>U_T$ and Mach cone dominance in the opposite limit.

\begin{figure}[t]
\begin{center}
\epsfig{width=11cm,figure=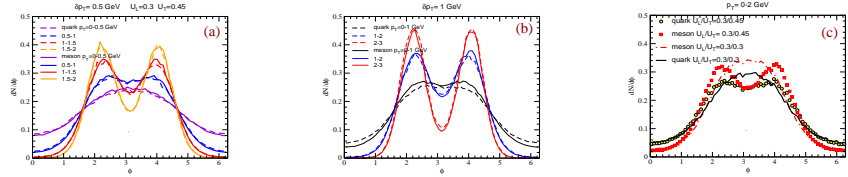}
\caption{\label{coalmc}(color online) Conical signal integrated over different momentum bins.  The amplitudes (close to the respective thermal weights) have been normalized so as to overlap}
\end{center}
\end{figure}

\begin{figure}[h]
\begin{center}
\epsfig{width=10cm,figure=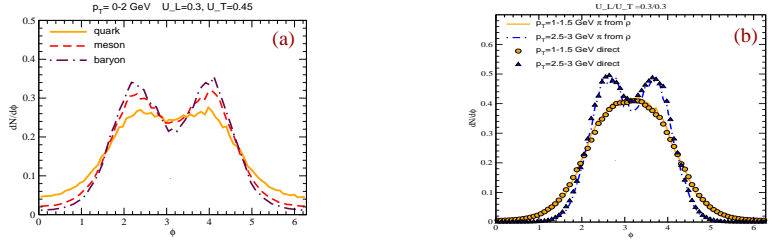}
\caption{\label{coalresbar}(color online) Conical signal integrated over different momentum bins. See Fig. \ref{coalmc} for normalization }
\end{center}
\end{figure}

As can be seen in Fig. \ref{coalmc}, coalescence can produce a peak in mesons where there was none in quarks within the same momentum regime {\em provided} the momentum  is integrated over large enough bins (Notice in Fig.2(b) that mesons at $p_T$ 1-2 GeV shows two peaks even
if they come from quark at 0-1 GeV that have only one peak).  
This is due to the signal sharpening by coalescence shown in panel (c) of Fig \ref{gausspeak}, and non-collinearity at low $p_T$.  The latter requirement causes a deviation from the exponential distribution,resulting in an hadron distribution that peaks at higher $p_T$ wrt to the quark one. Thus higher momenta contribution, where the conical signal is stronger, are more important with the effect being larger for baryons as shown in Fig.\ref{coalresbar}.  This effect disappears if momentum bins are narrow enough, and the relative importance of each bin is absorbed into the normalization. This can be seen in panels (a) and (b) of Figs  \ref{coalmc}: If $dN/d\phi$ is binned narrowly in momentum, the shape of the signal is the same before and after coalescence (and cones are more apparent at higher $p_T$).  
At high enough $p_T$ ($>2$ GeV, where the impact of unthermalized hard quarks is anyways non-negligible\cite{muellerpaper}) this effect also disappears, since coalescence is nearly collinear, and thermal and coalescing distributions become identically exponential.   
The Mach angle is independent of quark number in the hadron, since it reflects the partonic flow pattern.  The depth of the signal, however, scales with quark number.
This observation is so far compatible with experimental observations on particle-identified soft-hard correlations \cite{expbar}.
Panel (b) in Fig. \ref{coalresbar} confirms that resonance decays (in this case $\rho \rightarrow \pi \pi$) do not significantly modify the angular correlations.  Thus, the high admixture of resonances in the freezing out hadrons (which would naturally solve the entropy problem, typical of coalescence models \cite{Greco:2007nu,cassing}) does not significantly affect the Mach cone detectability.  Since the conical signal is largely due to higher-$p_T$ hadrons, where resonance decay products are collinear, this is not surprising.
In conclusion, we have shown that the conical signal can be significantly strengthened by quark coalescence, since it combines relatively low-momentum partons to higher momentum hadrons.  When the coalescing quarks have a constituent mass, this effect is not cancelled out by the weakening of the flow signal at low momentum.  Thus, when the conical signal is integrated in a wide enough transverse momentum bin, coalescence can produce a hadronic mach cone where thermal smearing was strong enough to quench the conical signal at quark level.

Before it becomes clearer to what extent this effect is responsible for 
the experimentally observed conical-looking signal in hard-soft 
correlations, more realistic comparisons including background transverse and elliptic flow must be performed.  By using experimental observables such as identified baryon/meson correlations \cite{expbar} and binning by the angle between the hard trigger and the reaction plane, coalescence and Cooper-Frye freeze-out scenarios could become experimentally distinguishable through such correlations measurements.\\

\noindent \textbf{Acknowledgements} \noindent J.N.\ and M.G.\
acknowledge support from DOE under Grant No.\ DE-FG02-93ER40764. 
G.T. was supported by the Helmholtz
International
Center for FAIR within the framework of the LOEWE program
(Landesoffensive zur Entwicklung Wissenschaftlich-\"Okonomischer
Exzellenz) launched by the State of Hesse.  We thank D. Molnar and R. Hwa for discussions.





\end{document}